# Oxygen Sensing Without Organic Molecules: Mixed-phase $TiO_2$ as Cost-effective Ultrasensitive Optical Sensors.

R. Rega[1], A. Fioravanti[2], F. Borbone[3], M. Mazzocchi[4] and S. Lettieri[5*]

[1] Institute of Applied Sciences and Intelligent Systems, National Research Council (CNR-ISASI), Via Campi Flegrei 34 - 80078 Pozzuoli, ITALY

[2] Institute of Sciences and Technologies for Sustainable Energy and Mobility, National Research Council (CNR-STEMS), Via Canal Bianco 28 – 44124 Ferrara, ITALY

[3] Department of Chemical Sciences, University of Napoli Federico II, Via Cupa Cintia 21 - 80126 Napoli, ITALY.

[4] Institute of Science, Technology and Sustainability for Ceramics, National Research Council (CNR-ISSMC), Via Granarolo 64 - 48018 Faenza, ITALY.

[5] Department of Physics "Ettore Pancini", University of Napoli Federico II, Via Cupa Cintia 26 - 80126 Napoli, ITALY.

*Corresponding author. E-mail: Stefano.lettieri@unina.it

**Abstract**. Oxygen ($O_2$) detection is commonly carried out via either fluorescence-based optical sensors or chemoresistive sensors. Each approach has its own limitations. Optical sensors require the molecular design and synthesis of specific organic fluorescent species which can be costly and less stable. Chemoresistive sensors, despite presenting advantages in using more cost-effective inorganic materials, are often limited by low sensitivities to $O_2$ at ppm concentrations and by their inability to operate at room temperature. In this work, we demonstrate the detection of $O_2$ at concentrations as low as a few tens of ppm at room temperature by using titanium dioxide ($TiO_2$) mixed-phase nanoparticles as optical sensors. By simultaneously measuring the photoluminescence of nanoparticles in rutile and anatase phase, $O_2$ detection was achieved in the concentration range of 30-500 ppm, with a response curve well-calibrated by a Langmuir function. Good response promptness and repeatability are also demonstrated. This approach to $O_2$ optical sensing offers two intrinsic advantages over the most commonly used methodologies: (1) use of a cost-effective, easy-to-prepare and stable of the sensitive material compared to those typically employed in optical sensing, and (2) improved room-temperature detection efficiency in the low $O_2$ concentration range, outperforming most commonly-used chemoresistive sensors.

## 1. Introduction

The discovery of molecular oxygen ($O_2$) by Carl Wilhelm Scheele in 1772 and, independently, by Joseph Priestley in 1774 sparked a chemical revolution and paved the way to new interest in the study of gases. $O_2$ has been essential for life for more than 2 billion years, ever since the first organisms able to transduce solar energy into the chemical energy of carbon bonds started to evolve [1]. Measuring $O_2$ concentration is of critical importance in a wide range of fields and applications, including environmental and marine sciences [2,3], gardening [4], food production [5,6], medical diagnostics [7,8] and many others. Several applications of $O_2$ sensors play an important role in the fields of biology and biochemistry, as discussed in several review papers [9–12].

Among the large-scale industrial applications, the most relevant use of $O_2$ sensors involves the analysis of exhaust gas of internal combustion engines. Monitoring $O_2$ concentrations in exhaust gases enables the control or air-to-fuel ratio, ensuing optimal combustion with reduced harmful emissions and improving fuel efficiency. Additionally, $O_2$ gas sensors are crucial for specific industrial process control, such as optimization of combustion efficiency in heating furnaces [13], and for ensuring occupational safety in work environments like mines and underground spaces, where the worker could be subject to a decrease in oxygen levels [14].

Optical sensors based on photoluminescence (PL) variations are widely employed for $O_2$ sensing. These devices measure changes in PL (intensity and/or lifetime) of specific chemical species caused by the reversible adsorption of $O_2$. The magnitude of these changes is used to quantify the concentration of $O_2$ concentration in the liquid or gaseous medium in which the sensor is used. Optical sensors offer peculiar advantages such as the ability to perform contact-free and non-invasive measurements, high accuracy, and remote sensing capabilities via the use of optical fibers. Comprehensive reviews on the topic can be found in the works by Bitting [2] and by Wang and Wolfbeis [9]. However, these sensors also present drawbacks. In particular, PL-based optical sensors usually employ the use of organic dyes and pigments specifically designed and optimized for this purpose, common examples being metalloporphyrins and transition metal polypyridyl complexes [15,16], with evident drawbacks related to fabrication costs.

The most well-established types of chemoresistive sensors are based on thick or thin films of metal oxide semiconductors (MOXS) used as sensing elements. The physical mechanisms underlying the chemoresistive detection of $O_2$ are well-known and extensively described, so they will not be detailed here. Unlike many organic materials commonly used in optical sensing, MOXS are inexpensive and straightforward to prepare. An analysis of the literature regarding $O_2$ sensing via chemoresistive sensors shows that most reports deal with detecting $O_2$ concentration ranging from approximately 0.1% to 20%, while reports of efficient detection at concentrations below 1000 ppm via chemoresistive devices are rare. Notably, the detection of $O_2$ at levels in the tenths of ppm would be relevant for various practical applications, such as in food and beverages industry [15].

Although MOXS are mostly used in chemoresistive devices, it is important to underline that some of them can also be employed as optical sensors, either in simple the form of nanoparticle thick or thin films [16–19] or as more complex photonic structures [20–22]. For example, gas-dependent PL has been demonstrated for $SnO_2$ [23–25], ZnO [26–28], undoped $TiO_2$ [29,30], rare earth doped $TiO_2$ [31,32], $Nb_2O_5$ [33] and nickel/cobalt titanates ($MTiO_3$, M = Ni and Co) [34]. These findings suggest interesting potential for $O_2$ optical sensing via MOXS nanoparticles, especially when considering two key points. (1): MOXS nanoparticles are stable and typically cheaper and easier to synthesize than the organic molecules commonly employed in optical oxygen sensing. (2): Differently from chemoresistors, PL-based sensors do not need thermal activation of charge carriers, so that in principle they can operate even at room temperature.

In this work, we demonstrate the possibility of detecting $O_2$ at concentration as low as a few tens of ppm at room temperature by exploiting the PL of $TiO_2$ nanoparticles in mixed phase, i.e., nanoparticles partially converted from the anatase to the rutile phase. We present evidence of $O_2$ detection in the range of 20 - 500 ppm, with a response curve that can be calibrated using a Langmuir function. Good promptness and repeatability of the responses are also demonstrated. The following section outlines the methodology and procedures employed. Subsequently, the experimental results are presented, discussed and compared with those typically obtained under similar conditions (i.e., $O_2$ concentrations and room temperature) using standard chemoresistive devices. This comparison highlights the effectiveness of our straightforward approach.

## 2. Methods and procedures

### 2.1 Nanoparticle synthesis

$TiO_2$ nanopowders have been synthesized via wet-chemical synthesis. The synthesis procedure has been described in more details in previous articles [35,36]. Here we summarize the main steps: a solution of Ti(IV) n-butoxide in pure ethanol was slowly added under stirring to a 1:1 volume ratio ethanol/water solution. After stirring, the suspension was filtered by gravity, washed with water and diethyl ether and then dried overnight at a temperature of 100°C. Next, the powders were calcinated for 2h at different temperatures to obtain $TiO_2$ nanoparticles. For the present work we employed calcination temperatures of 500 °C, 650°C and 700 °C. The resulting samples are hereafter named as S500, S650 and S700, respectively. Anatase-to-rutile phase transition in $TiO_2$ typically occurs at temperatures of about 600 °C. Hence, we chose the calcination temperatures with the intention of obtaining one sample in pure anatase phase (S500) and two in mixed phase (i.e., showing both anatase and rutile phases) and possibly exhibiting differences in the size distribution of the nanoparticles.

### 2.2 X-ray diffractometry and SEM

Crystalline structures were examined via X-ray diffraction (XRD) by means of a Bruker D8 Advance diffractometer (Cu K radiation, 40 kV, 40 mA) with LYNXEYE detector. All XRD patterns (see Figure 1) were collected over the range 2θ = 10°–90°, with steps of 0.02° and 2 s of dwell time. The cell parameters, phase percentage and Bragg R factors were obtained by Rietveld analysis software FullProf (release 2021) [37]. Average crystallite sizes were calculated with the Williamson-Hall method [38].

SEM images of the nanoparticles were obtained via a field emission gun SEM (FEG−SEM) FEI/Thermo Fisher Nova NanoSem 450. The samples were previously sputtered with a nanometric layer of an Au/Pd alloy using a Denton Vacuum Desk V TSC coating system. The size distribution of the nanoparticles was determined from the SEM images by extracting the Feret diameter of the contour of each particle through the ImageJ software. The distributions were plotted as a histogram of the frequency with a bin of 5 nm.

### 2.3 Photoluminescence measurements

PL measurements of the $TiO_2$ nanoparticles have been performed by means of a home-built setup which allows to measure the PL spectra at controlled values of excitation wavelength and of $O_2$ concentration. The samples are placed in a home-built test chamber equipped with a fused silica entrance optical window. $O_2$ concentration was controlled by mixing the gas flows from a $N_2$ cylinder (99.9995% purity) and certified cylinders containing $O_2$ at concentration of 100 ppm and 5000 ppm.

Optical excitation was provided by a Xe lamp coupled with a double-grating excitation monochromator. A system of converging lens and optical filters was employed to image the illuminated part of the sample on the input slit of a monochromator coupled with a Peltier-cooled CCD camera. PL intensity data are reported in cps (count-per-seconds) units, indicating the CCD counts of the digitally processed PL intensity divided by the illumination time, which amounted to 3 s for each data point. The entire system was controlled via a computer using a home-made LabView program.

Illumination was provided at an excitation wavelength of 370 nm. Most measurements were conducted using a method where the nanoparticles were illuminated only during PL acquisition, a method referred to as “targeted illumination”. This was achieved by blocking the excitation light with a mechanical shutter synchronized to the CCD camera. A measurement method in which the sample was under continuous exposure to the excitation light, referred to as “uninterrupted illumination” was also tested. All measurements were performed at room temperature.

## 3. Results and Discussion

### 3.1 Experimental results

The XRD diffractograms of the investigated samples are shown in Figure 1. As expected, it is seen that sample S500 exhibited only the anatase phase (space group $I4_1/amd$, JCPDS File No. 21-1272), while the other two samples showed the presence of both anatase and rutile (space group $P4_2/mnm$, JCPDS File No. 21-1276) phases, without extra phases present. The structural parameters obtained from the XRD data are reported in Table 1. The data analysis by Rietveld method indicates that the calcination at 700°C provides slightly larger rutile nanoparticles than at 650°C. The Bragg R-factor for the rutile phase is larger in S700 than S650, indicating a more accurate goodness-of-fit of the Rietveld analysis in S650. Nonetheless, SEM analysis shown later also supports the conclusion that slightly larger nanoparticles are obtained via 700°C calcination.

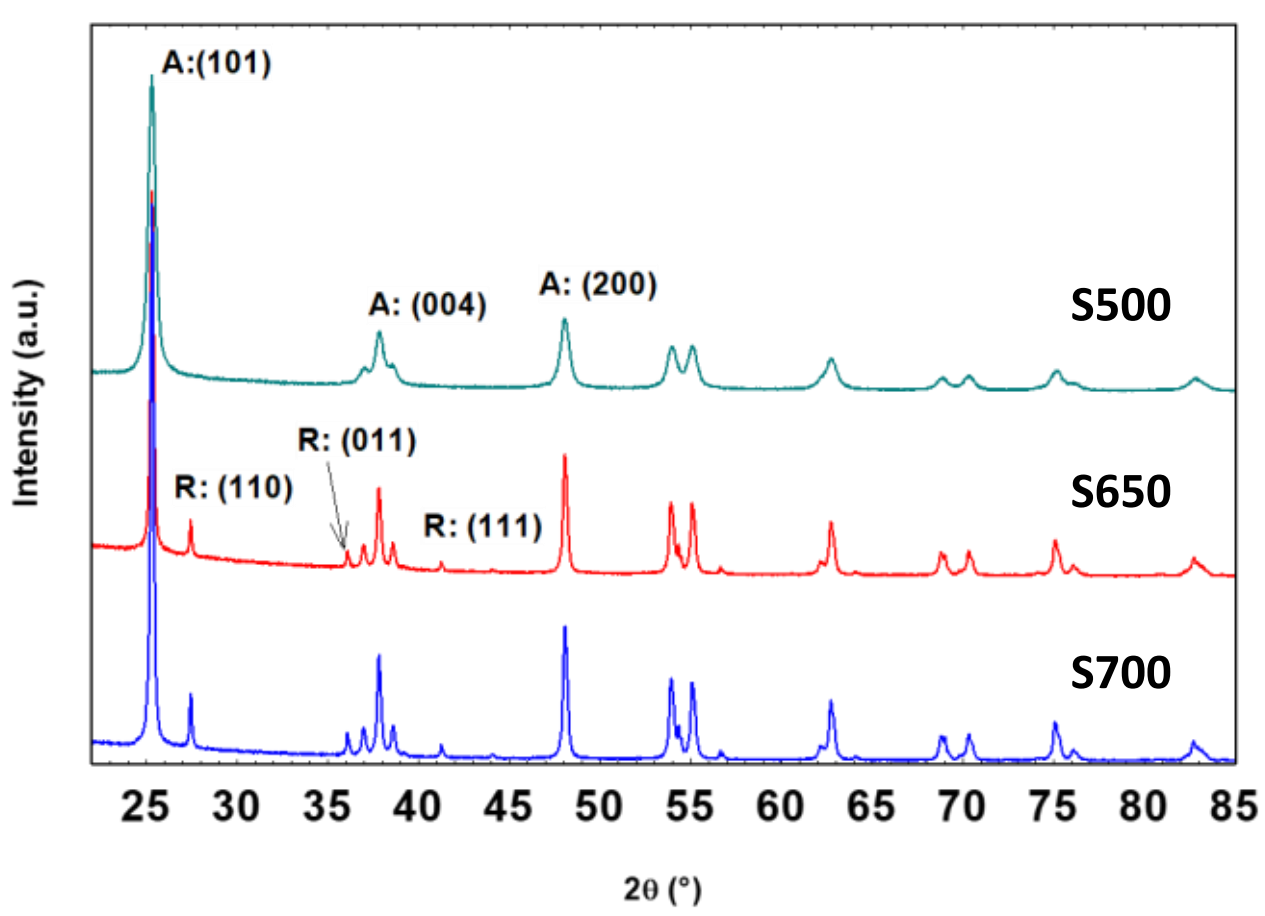

**Figure 1:** X-ray diffractograms for S500, S650 and S700 samples. The labels identify diffraction peaks corresponding to different crystalline planes of anatase (A) and rutile (R).

| | Anatase | | | | | Rutile | | | | |
|---|---|---|---|---|---|---|---|---|---|---|
| *Sample* | $a$ (Å) | $c$ (Å) | % | $R_B$ | $av.size$ $(nm)$ | $a$ (Å) | $c$ (Å) | % | $R_B$ | $av.size$ $(nm)$ |
| S500 | 3.78455(6) | 9.5025(5) | 100 | 4.38 | 40 | - | - | 0 | - | - |
| S650 | 3.78414(6) | 9.5109(2) | 90 | 8.13 | 40 | 4.5933(2) | 2.9580(2) | 10 | 8.45 | 50 |
| S700 | 3.78433(5) | 9.5108(2) | 90 | 7.45 | 50 | 4.5933(2) | 2.9581(2) | 10 | 13.2 | 70 |

**Table 1**: Values of the cell parameters (a, c), phase percentages, Bragg R factors and average nanoparticle sizes for the investigated samples.

Figure 2 shows representative SEM images of the nanoparticle samples, alongside histograms of the Feret diameter distributions. The average and standard deviation of the size distributions were determined to be 40±16 nm for S500, 52±24 nm for S650 and 55±23 for S700. Consistent with the data from the XRD measurements, it is thus observed that higher calcination temperature leads to slightly larger nanoparticles.

A direct comparison between the size histograms of S650 and S700 shows that the latter contains a higher proportion of larger nanoparticles (approx. 80 nm and more), resulting in the larger average size. This agrees with the indication from XRD, which indicates that the larger average size in S700 is mainly due to the comparatively larger growth of rutile nanoparticles compared to S650.

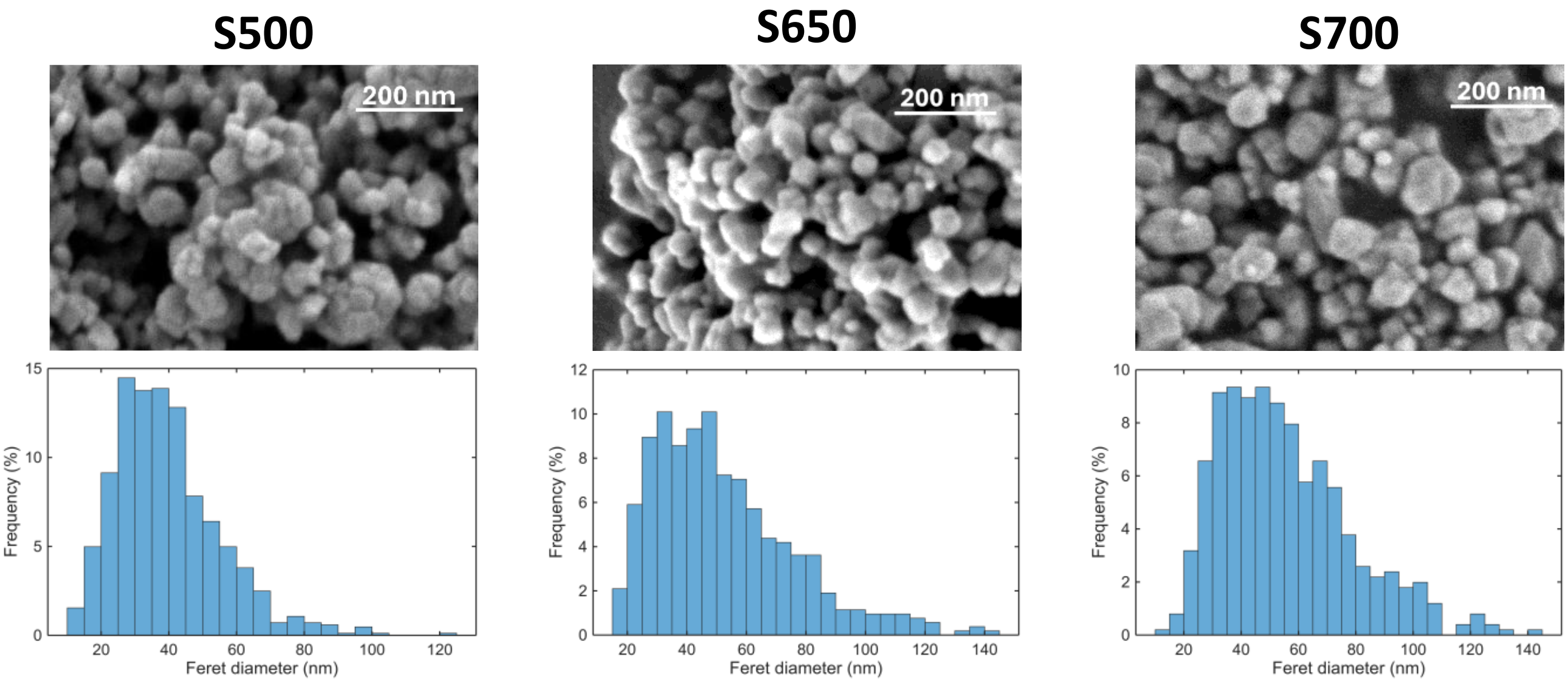

**Figure 2**: SEM images and corresponding histograms of Feret diameters of the nanoparticles for samples S500 (left), S650 (center) and S7500 (right).

Dynamical PL responses of mixed-phase $TiO_2$ toward $O_2$ were determined according to the procedure described previously. Mixed-phase samples containing both anatase-phase and rutile-phase $TiO_2$ nanoparticles were chosen because, as documented in the literature [39,40], the two polymorphs exhibit PL spectra whose intensity changes in opposite manner as the material interacts with $O_2$. Anatase $TiO_2$ luminesces in the visible range (approx. 450-700 nm wavelength interval, PL maximum at about 520 nm), and its PL intensity decreases when the material is exposed to $O_2$. On the other hand, rutile $TiO_2$ luminesces in the near-infrared range (approx. 780–920 nm wavelength interval, PL maximum at about 840 nm), and its PL intensity is increased by $O_2$ exposures. Additionally, the spectral separation of these two PL bands allows them to be easily distinguished and simultaneously detected in samples containing both crystalline phases.

This phase-dependent behavior of PL intensity highlights mixed-phase $TiO_2$ as a potentially efficient optical sensor, through the simultaneous detection the PL spectra originating from the two polymorphs and through the use of their ratio (PL intensity of rutile divided by PL intensity of anatase) as the optical signal for the sensing transduction. As discussed in a previous work [41], this approach leads to sensor responses which are intrinsically larger than the one obtainable from single-phase samples.

Figure 3 shows the results obtained by exposing the samples to different concentrations of $O_2$ for 5 minutes, followed by recovery in pure $N_2$ for 10 minutes. The PL intensity spectra recorded over time are reported as contour color plots in Fig. 3(A) for sample S700 and in Fig. 3(B) for sample S500. The green horizontal lines indicate the time instant when the pure $N_2$ flow was introduced in the experimental chamber ("0 ppm $O_2$"), while the white lines mark the moments when $O_2$ was introduced at different concentrations.

The stabilized PL spectra recorded in pure nitrogen (black curve) and after 5 minutes of exposure to 100 ppm $O_2$ (red curve) are shown for samples S700 in Fig. 3(C) and for S500 in Fig. 3(D). The colored bands highlight the wavelength intervals used to calculate the total PL signals for visible and near-infrared PL, as explained next. The previously described behaviors of the two emission bands in the mixed-phase sample S700 are clearly observed. The PL emission band at 800-900 nm is less intense than the broad visible band, but the increase of its intensity during exposure to $O_2$ is clearly noticeable in Fig. F1(A). Instead, as expected, only a visible PL band was observed in the anatase S500 sample.

For the remainder of this work, we focus solely on mixed-phase samples and we obtained, for each of these samples, three different experimental responses indicated as $R_{vis}, R_{nir}$ and $R_{tot}$. The first two are extracted from the experimental data relative to visible (hence "vis") and near-infrared ("nir") PL components. They are defined as follows:

$$R_{vis} = I_{0,vis}/I_{vis} \quad ; \quad R_{nir} = I_{nir}/I_{0,nir}$$

Here, the "vis" and "nir" total PL intensities are obtained by numerical integration with respect to the wavelength $\lambda$ over the intervals $\Delta\lambda = 500 - 650\ nm$ and $\Delta\lambda = 800 - 900\ nm$. These spectral intervals are evidenced in Fig. 3(C) and Fig. 3(D) by the colored background. The subscript "0" refers to the measured intensity in absence of analyte gas (i.e., in flowing $N_2$).

The "total response" $R_{tot}$ exploits the aforementioned idea of using the ratios $(I_{nir}/I_{vis})$ as the transduction signal, so that $R_{tot} = (I_{nir}/I_{vis})/(I_{0,nir}/I_{0,vis})$. It is immediate to see that this is equivalent to defining it as the product of the two previous responses, which is $R_{tot} = R_{vis} \cdot R_{nir}$.

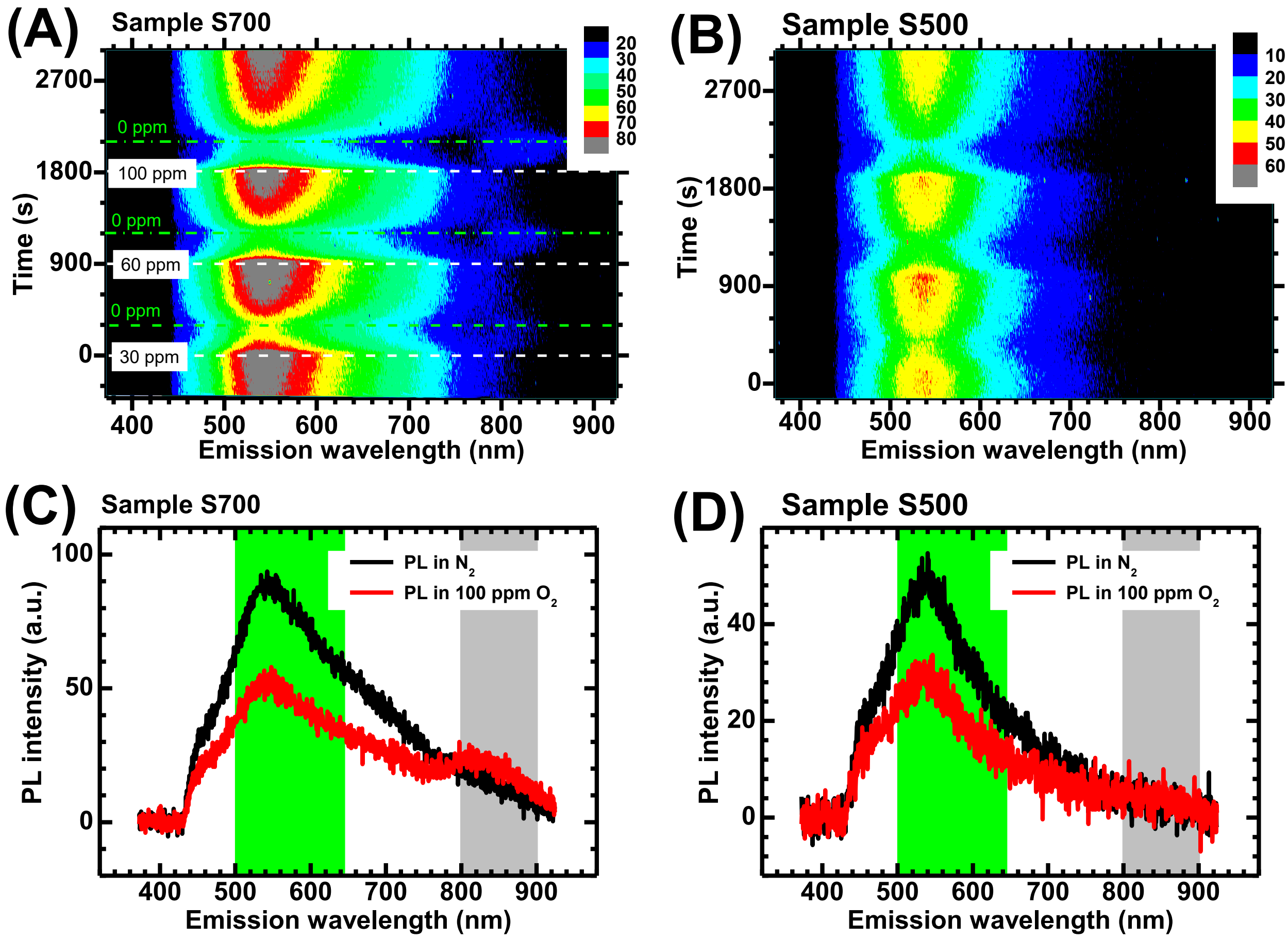

**Figure 3**. (A) Dynamical response of PL intensity for S700 sample to exposure to $O_2$ at concentrations of 30, 60 and 100 ppm. The horizontal dashed lines mark the instant when the composition of the gas flow is changed. (B) Same for the S500 sample. To improve the readability, the time instants when the gas flows are activated are not marked. (C): Stabilized PL spectrum of S700 in $N_2$ (black curve) and 100 ppm $O_2$ (red curve). (D) Same for S500 sample.

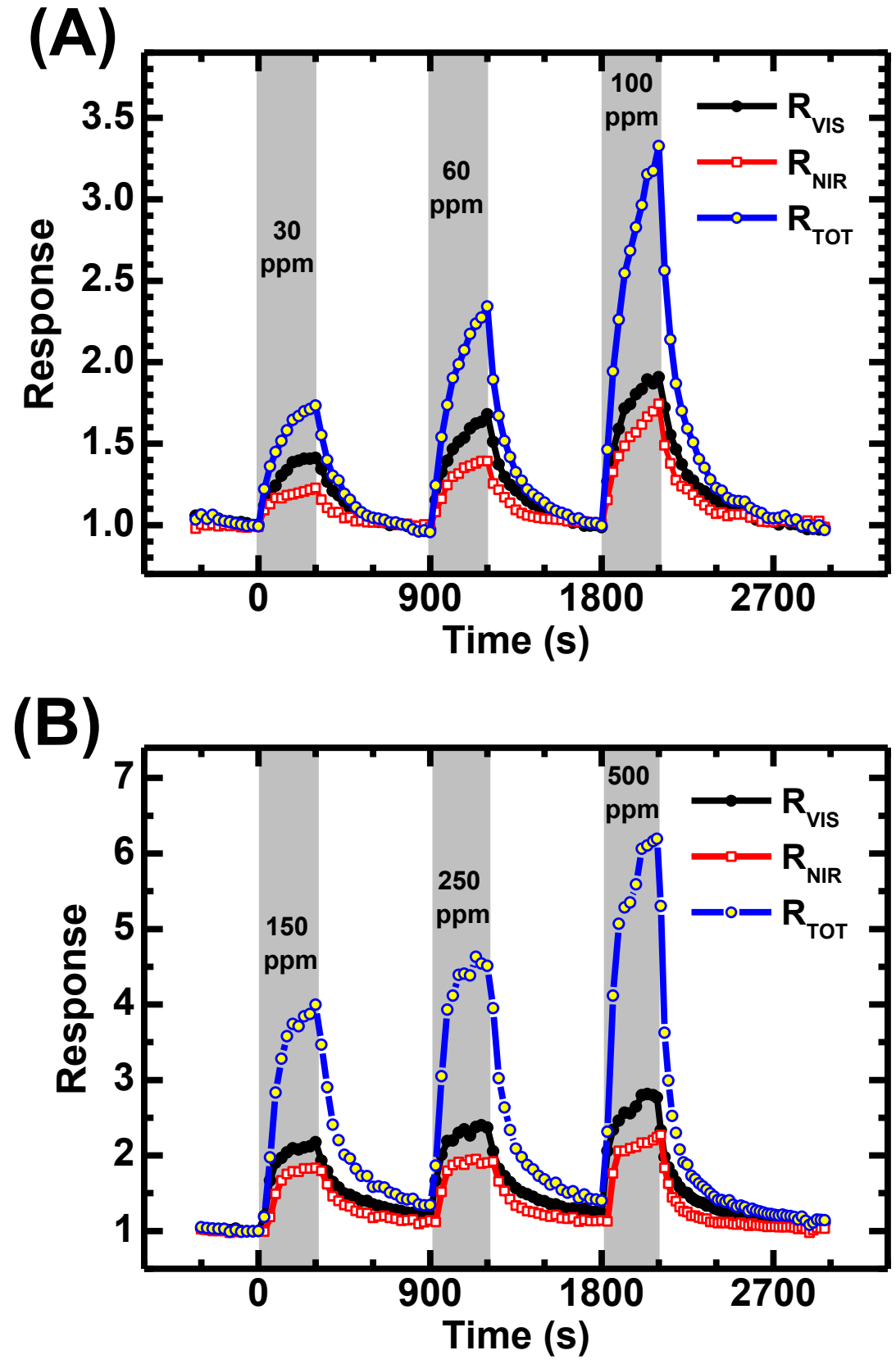

**Figure 4**: (A) Dynamic response of S700 sample toward $O_2$ at concentration of 30, 60 and 100 ppm. (B): Same for $O_2$ at concentration of 150, 250 and 500 ppm. The black, red and blue curve refer to the visible, near-infrared and total responses ($R_{VIS}$, $R_{NIR}$ and $R_{TOT}$) as defined in the text.

Figure 4 shows the dynamical responses $R_{VIS}$ (black curve), $R_{NIR}$ (red curve) and $R_{TOT}$ (blue curve) exhibited by the S700 sample in response to $O_2$ varying in the range from 30 to 500 ppm. Specifically, Figure 4(A) shows responses to $O_2$ concentrations of 30, 60 and 100 ppm, while Figure 4(B) shows responses to $O_2$ concentrations of 150, 250 and 500 ppm. In each instance, the sample was exposed to $O_2$ for 5 minutes, followed by 10 minutes of recovery under pure $N_2$ flow.

From the dynamical response, we calculated the response rise time and fall time (i.e., recovery time), defined as the time required for the response to reach 90% of its maximum amplitude during the $O_2$ exposure and the time needed to drop to 10% of its maximum amplitude during the exposure to $N_2$, as shown in Figure 5(A).

The characteristic times obtained from the different experimental responses are reported in Figure 5(B) (rise times) and Figure 5(C) (fall times) in $10^2$ s units. It is important to recall that the experiments were performed with data acquisition at intervals of 30 seconds, introducing an intrinsic sensitivity uncertainty in estimations of response times. The results indicate that the VIS response is faster than the NIR response and that the rise times are slightly shorter than the fall times. To help comparison, the histograms in Figures 5(B) and 5(C) are presented using the same scale.

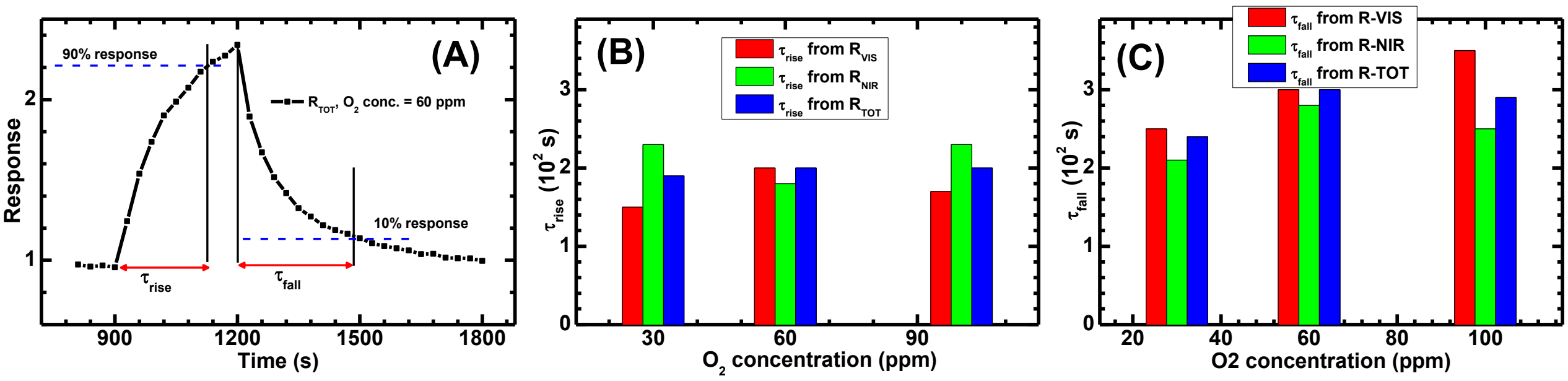

**Figure 5**. (A): Dynamical response $R_{TOT}$ measured for $O_2$ concentration of 60 ppm: the plot illustrates how the rise and fall times are calculated from the data. (B): Rise times of the responses $R_{VIS}, R_{NIR}$ and $R_{TOT}$ at $O_2$ concentration of 30, 60 and 100 ppm. (C): Fall times, same as for plot (B). All the data refer to the PL response of sample S700.

To evaluate response repeatability, successive exposures to identical $O_2$ concentrations were tested, using the same time intervals for the exposure and recovery as in Fig. 4. An example is presented in Figure 6, using the same color scheme as in Figure 4, showing successive exposures 80 ppm $O_2$. The results indicate satisfactory repeatability.

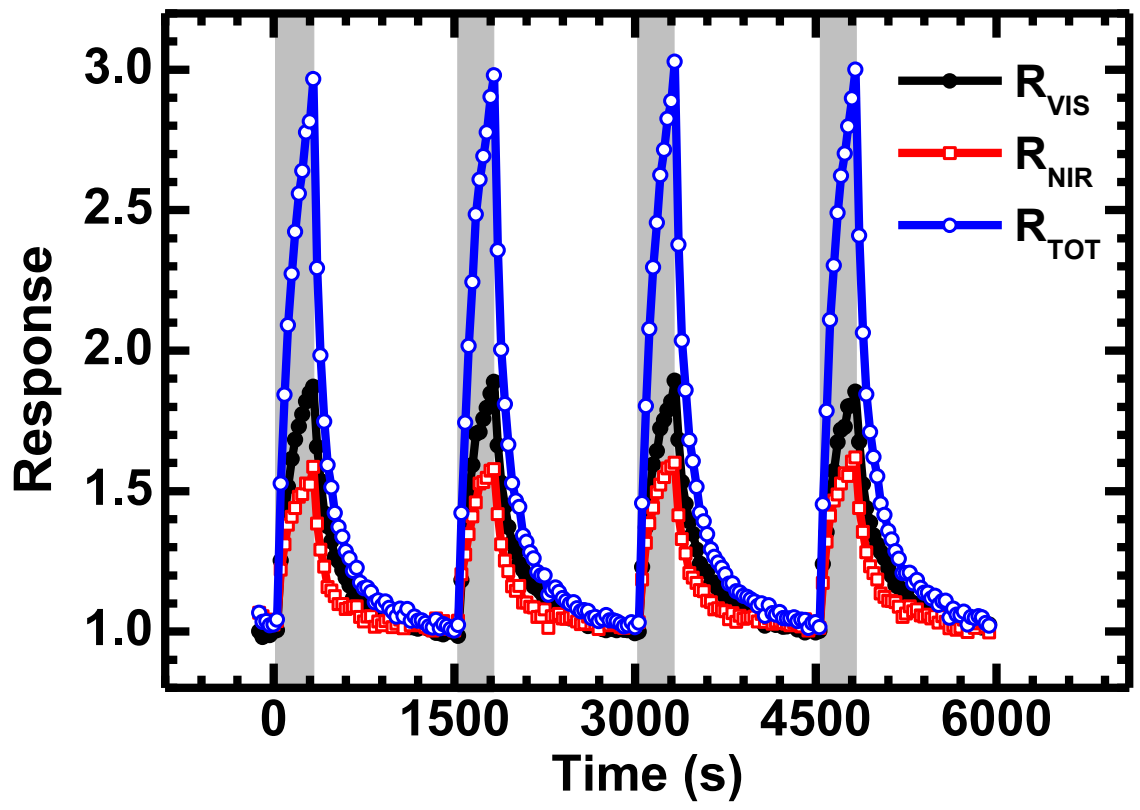

**Figure 6**. Dynamic PL responses of S700 sample toward repeated exposures to $O_2$ at 80 ppm concentration. The black, red and blue curve refer to the visible, near-infrared and total responses ($R_{VIS}, R_{NIR}$ and $R_{TOT}$) as defined in the text.

The experimental results for S650 nanoparticles closely resemble those for S700, with the only noticeable difference being slightly lower values response values for S650 at larger $O_2$ concentrations. The response data obtained for S700 and S650 are presented in Table 2, reporting the maximum values for $R_{VIS}, R_{NIR}$ and $R_{TOT}$ at the different $O_2$ concentration tested for the two nanopowders. The Table also includes the rise and fall times extracted from the dynamical $R_{TOT}$ response.

The maximum responses versus $O_2$ concentrations are summarized in Figure 7. Fig. 7(A) displays the responses obtained for sample S700, while Fig. 7(B) compares the $R_{TOT}$ of S700 (empty circles) and S650 (black triangles) as a function of $O_2$ concentration. The responses are well-fitted by a Langmuir function in the form $R(C) = 1 + aC/(1 + bC)$, where C is the $O_2$ concentration. This function allows calibration of the responses through data fitting, accounting for $R = 1$ in absence of $O_2$ and for the plateau of the response at high $O_2$ concentrations. The good fitting provided by the Langmuir-type function strongly suggests that the $O_2$ molecules, within the investigated concentration range, adsorb at the semiconductor surface according to a Langmuir kinetics and that the PL intensity changes scales are directly proportional to the surface density of adsorbed $O_2$ molecules [42,43].

| | Sample S700 | | | | | Sample S650 | | | | |
|---|---|---|---|---|---|---|---|---|---|---|
| $[O_2]$ $(ppm)$ | $R_{VIS}$ | $R_{NIR}$ | $R_{TOT}$ | $\tau_{rise}$ $(10^2\ s)$ | $\tau_{fall}$ $(10^2\ s)$ | $R_{VIS}$ | $R_{NIR}$ | $R_{TOT}$ | $\tau_{rise}$ $(10^2\ s)$ | $\tau_{fall}$ $(10^2\ s)$ |
| 30 | 1.4 | 1.2 | 1.7 | 1.9 | 2.4 | 1.3 | 1.1 | 1.4 | 2.1 | 2.4 |
| 60 | 1.7 | 1.4 | 2.3 | 2.0 | 3.0 | 1.6 | 1.2 | 1.9 | 2.7 | 2.7 |
| 80 | 1.8 | 1.6 | 2.9 | 1.9 | 2.9 | - | - | - | - | - |
| 100 | 1.9 | 1.7 | 3.3 | 2.0 | 2.9 | 2.0 | 1.4 | 2.8 | 2.4 | 3.3 |
| 150 | 2.2 | 1.8 | 4.0 | 1.7 | 3.0 | 2.1 | 1.8 | 3.8 | 2.0 | 3.1 |
| 250 | 2.4 | 1.9 | 4.6 | 1.8 | 3.0 | 2.2 | 2.0 | 4.4 | 1.8 | 2.7 |
| 500 | 2.8 | 2.3 | 6.3 | 1.8 | 2.4 | 2.3 | 2.4 | 5.5 | 1.6 | 2.6 |

**Table 2.** Responses and characteristic response times for samples S700 and S650. The rise time and fall time are calculated from the RTOT response curves (i.e., blue curves in Figure 4).

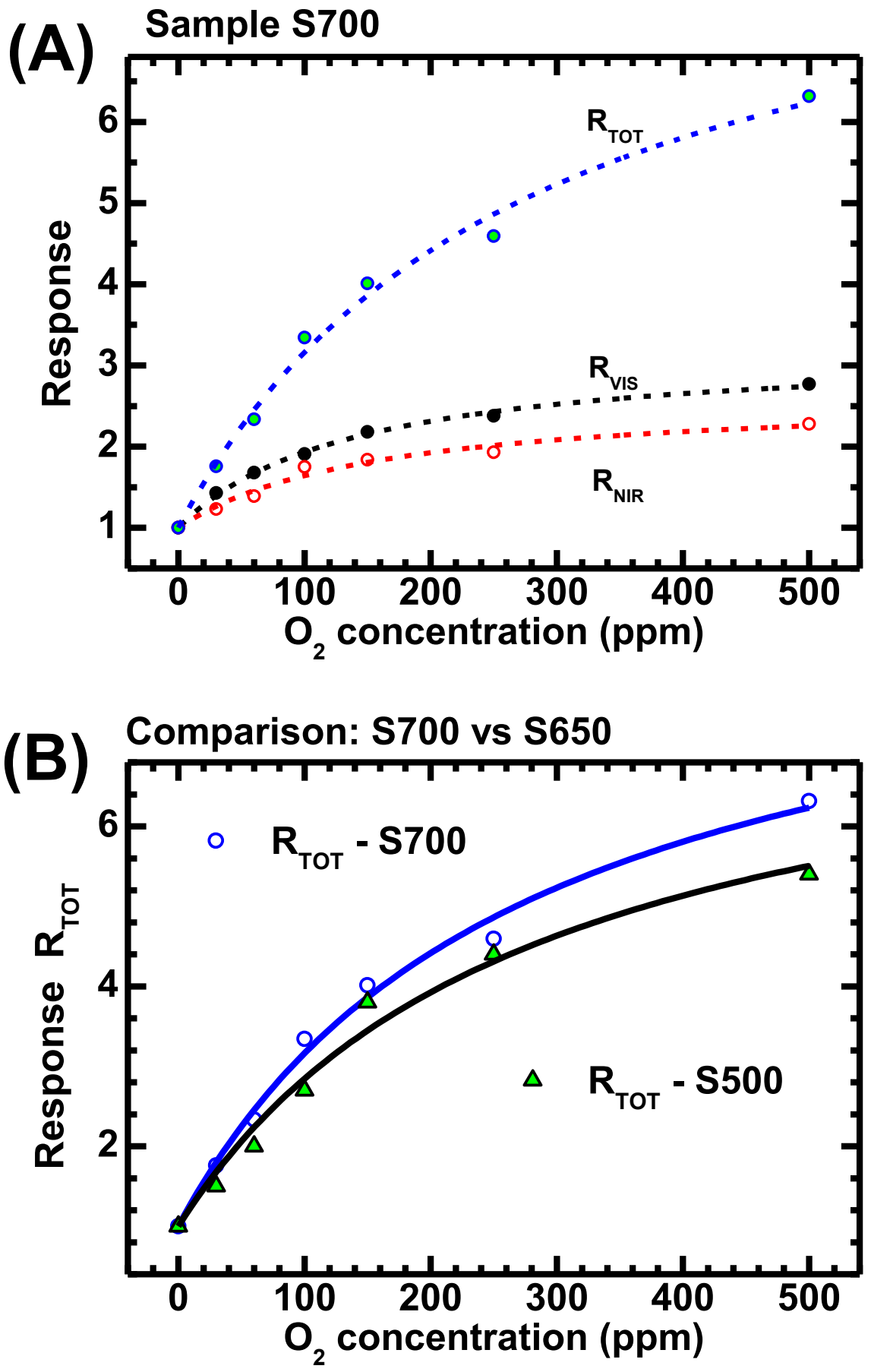

**Figure 7.** (A): The data points indicate the responses $R_{VIS}$, $R_{NIR}$ and $R_{TOT}$ obtained for sample S700 plotted vs $O_2$ concentration. The curves are a best-fit of the data using the Langmuir function defined in the text. (B) Comparison of the $R_{TOT}$ response curves obtained for S700 and S650. In order to improve the readability of the figure, only the $R_{TOT}$ data are reported. The curves again represent the best-fit of the data using the Langmuir function.

The data presented thus far were collected in "targeted illumination" conditions, where the samples were illuminated only during the PL acquisition (3 seconds of illumination each 30 seconds). Given the potential issues that might be caused by residual impurities (which is one of the positive side effects of using high temperatures in chemoresistors)

Measurements were also performed in the "uninterrupted illumination" configuration by removing the shutter. In this configuration, the $TiO_2$ nanoparticles remained illuminated by the UV excitation light for the entire observation period. The motivation for such a test was to check if the rise of sample temperature caused by prolonged UV absorption would favor the desorption of eventual residual impurities and improve the measurement repeatability. The evidence shows that this is not the case: continuous illumination caused background signal drift and degraded responses quality. An example is illustrated in Figure 8, showing a comparison of the dynamical $R_{VIS}$ response of S700 when subjected to targeted illumination (red curve, the same data as in Figure 4) versus uninterrupted illumination (black curve). It is evident that under continuous illumination the PL intensity in $N_2$ drifts over time (dashed black line), while targeted illumination minimizes such instability (dashed red line) and provides much more reliable results.

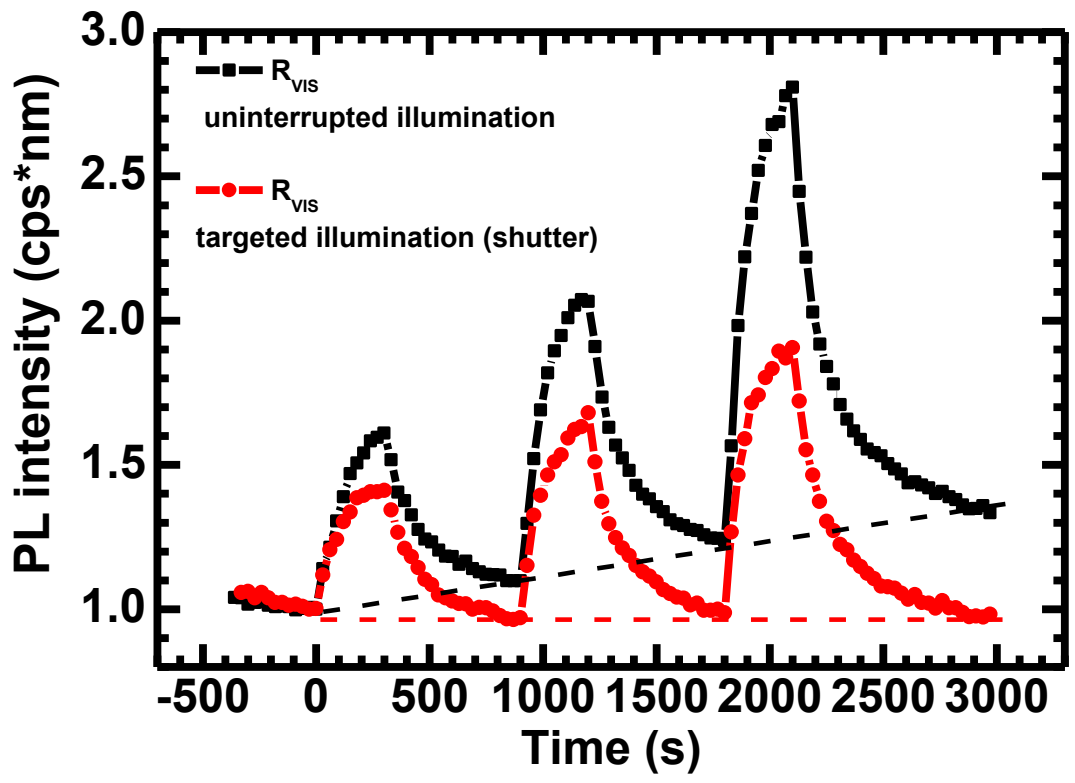

**Figure 8**. Black curve: dynamical response to $O_2$ for samples S700 obtained for uninterrupted illumination (black curve). Red curve: same but using targeted illumination (i.e., sample illuminated only during the data acquisition by use of a synchronized shutter). The black dashed lines evidence the drift of the baseline signal, which is absent for targeted measurements (red dashed line).

### 3.2 Discussion

The results presented in the previous section display two major characteristics, namely: (1). They are performed at room temperatures, and (2). Sensing of low concentrations (500 ppm and below) of $O_2$ is achieved. We discuss here the relevance of these two characteristics.

As outlined in the introduction, MOX-based chemoresistive sensors typically operate at temperature significantly above room temperature. Indeed, there are several practical factors that force to heat a chemoresistive sensor. Primarily, high temperatures are often needed to inject, via thermal activation, free carriers in the conduction band in n-type semiconductors (e.g. $SnO_2$) in order to obtain appreciably measurable values of electrical conductivity. Higher temperatures also accelerate the adsorption/desorption kinetics of the analyte gas and might favor the desorption of eventual surface contaminants, hence leading to faster and more stable sensor responses. High temperatures are also correlated with the efficiency of the chemoresistive process itself, which arises from $O_2$ molecules adsorbed at the surface extracting electrons from the conduction band and trapping them at the surface to form chemisorbed ions such as $O_2^-$, $O^{-2}$ or $O^-$. The latter species is considered to be the most reactive and prone to further reaction with reducing species if the semiconductor surface is in a temperature range of 300-450°C, which is in fact the working temperature for most metal oxide gas sensors [44]. For these reasons, the vast majority of scientific works regarding $O_2$ sensing via chemoresistive sensors refer to devices heated to temperatures usually in the range of 300-500°C (or more).

Additionally, the literature on $O_2$ detection at low concentrations via chemoresistors is quite limited. Most of the published studies report on $O_2$ detection at concentrations ranging between 1% and 100%. Table 3 summarizes data from the few relevant references we found reporting about $O_2$ concentrations below 1000 ppm. Notably, it is worth noting that the vast majority of the works in which the chemoresistive sensor operates at room temperature employ UV illumination to produce free carriers and measurable currents.

| Material | Temperature (°C) | Response to $O_2$ | Ref. |
|---|---|---|---|
| $TiO_2$/graphene | RT | $\Delta R/R_0 \approx 0.05$ at 134 ppm | [45] |
| LaOCl-doped SnO2 | RT, UV illumination | $R_0/R = 2.2$ at 250 ppm | [46] |
| Cr-doped $TiO_2$ nanowires | RT, UV illumination | $\Delta R/R_0 \approx 0.95$ at 1000 ppm | [47] |
| Undoped ZnO nanorods | RT | $\Delta R/R_0 \approx 0.8$ at 15 ppm | [48] |
| Mn-doped (3% wt.) ZnO nanorods | RT | $\Delta R/R_0 \approx 2$ at 15 ppm | |
| $Bi_2O_2Se$ microstructures | 100°C | $\Delta R/R_0 \approx 0.05$ at 100 ppm | [49] |
| $TiO_2$, thick film | 500°C | $\Delta R/R_0 \approx 0.25$ at 20 ppm | [50] |
| Nb-doped $TiO_2$, thick film | 500°C | $\Delta R/R_0 \approx 2.0$ at 20 ppm | |
| $TiO_2$, thick film | 116°C | $R/R_0 \approx 3.8$ at 100 ppm | [51] |
| P-doped (0.2% mol) $TiO_2$, thick film | 116 °C | $R/R_0 \approx 7.5$ at 100 ppm | |
| Undoped $SnO_2$, thin film | 250 °C | $\Delta R/R_0 \approx 0.015$ at 500 ppm | [52] |
| Pt-doped (0.3 % wt.) $SnO_2$, thin film | 250 °C | $\Delta R/R_0 \approx 0.045$ at 500 ppm | |
| ZnO nanowires on carbon microfibers | 200°C | $R/R_0 \approx 1.5$ at 500 ppm | [53] |
| | 250 °C | $R/R_0 \approx 1.5$ at 500 ppm | |
| $Nb_2O_5$, thick film | 200 °C | $R/R_0 \approx 1.2$ at 400 ppm | [54] |

**Table 3**. Literature data for chemoresistive sensors employed for detection of low (< 1000 ppm) concentrations of $O_2$ in inert background gas. operating at room temperature for O2 detection. The responses shown in the Table are typically defined as $(R_g - R_0)/R_0$ or as $R_g/R_0$, where $R_g$ and $R_0$ are the resistance measured in O2 and in inert background gas (typically N2), respectively.

The data in Table 3 highlights the advantages of employing optical sensors of $O_2$ based on the approach presented in this study. Firstly, PL-based sensing does not require thermal activation since the free charge carriers are generated via light absorption. Furthermore, using mixed-phase $TiO_2$ enables effective room-temperature detection of $O_2$ at concentrations of 100 ppm or less. For instance, the $R_{TOT}$ responses obtained in the present work (for example, the value 6.3 obtained for S700 at 500 ppm $O_2$ concentration) exceeds most values reported in Table 3. It is important to underline that these results were obtained using a material that is much simpler and cheaper to produce than the most commonly used optical oxygen indicators for optochemical sensors. Among these, representative and widely used examples include platinum-based porphyrins, such as PtOEP (CAS Number: 31248-39-2) and PtTFPP (CAS Number: 109781-47-7), which typically cost between \$500 and \$1000 per gram, or the ruthenium coordination complex $[Ru(BPY)_3]Cl_2$ (CAS: 14323-06-9), which is typically priced around \$200 per gram. In contrast, mixed-phase $TiO_2$ nanoparticles, such as the commonly available Aeroxide® P25 from Evonik are generally priced at just one or few dollars per gram at >99.5% purity.

## 4. Conclusions

This study demonstrates that the dual-response PL activity of $TiO_2$ nanoparticles in a mixed-phase form can be used as sensitive indicator of $O_2$ concentrations, harnessing the PL signals from the two phases to enable specialized room-temperature sensors for $O_2$ with performances not obtainable with traditional thick-film chemoresistive sensors.

By employing the ratio of PL intensities $(I_{nir}/I_{vis})$ as the transduction signal, we show that $O_2$ detection levels at room temperature can be achieved the lower ppm (30-500) range, with responses $R_{TOT}$ calibrated according to a Langmuir function and ranging from 1 to 6 (or, equivalently, $\Delta R/R_0 = 0 - 500\%$) with good repeatability. According to the literature, these results exceed the sensitivity of most standard chemoresistive sensors, even those operating at high temperatures.

A critical finding is highlighted regarding the illumination conditions employed during measurements, showing that continuous exposure to UV light during PL readings introduces greater instability in the responses, emphasizing the intricate balance between excitation conditions and sensor performance. A targeted procedure minimizing the illumination time, on the other hand, minimizes the baseline drift and enhances response reliability.

The optical sensing methodology explored here investigated here offers intrinsic advantages over both of the most commonly employed methodologies for $O_2$ sensing, as it leverages: (1) greater stability and lower production costs of the sensitive material compared to those typically employed in optical (PL-based) sensing, and (2) improved detection efficiency for low $O_2$ concentrations even at room temperature, as opposed to standard chemoresistive devices. Hence, we argue that these characteristics position mixed-phase $TiO_2$ nanoparticle-based sensors as promising candidates for various applications centered on $O_2$ detection.